\documentclass[10pt]{revtex4}
\usepackage{amsfonts}
\usepackage{amssymb,epsf}
\usepackage{latexsym}

\begin{document}

\title{Agegraphic Chaplygin gas model of dark energy}

\author{A. Sheykhi \footnote{
sheykhi@mail.uk.ac.ir}}
\address{Department of Physics, Shahid Bahonar University, P.O. Box 76175, Kerman, Iran\\
         Research Institute for Astronomy and Astrophysics of Maragha (RIAAM), Maragha,
         Iran}

\begin{abstract}
We establish a connection between the agegraphic models of dark
energy and Chaplygin gas energy density in non-flat universe. We
reconstruct the potential of the agegraphic scalar field as well
as the dynamics of the scalar field according to the evolution of
the agegraphic dark energy. We also extend our study to the
interacting agegraphic generalized Chaplygin gas dark energy
model.

\end{abstract}
\maketitle
\section{Introduction\label{Int}}
Over the course of the past decade, evidence for the most striking
result in modern cosmology has been steadily growing, namely the
existence of an exotic  dark energy component which has negative
pressure and pushes the universe to accelerated expansion
\cite{Rie}. Of course, a natural explanation to the accelerated
expansion is due to a positive tiny cosmological constant. Though,
it suffers the so-called \textit{fine-tuning} and \textit{cosmic
coincidence} problems. A great variety of scenarios have been
proposed to explain this acceleration while most of them cannot
explain all the features of universe or they have so many
parameters that makes them difficult to fit. For a recent review
on dark energy proposals see \cite{Pad}. Many theoretical attempts
toward understanding the dark energy problem are focused to shed
light on it in the framework of a fundamental theory such as
string theory or quantum gravity. Although a complete theory of
quantum gravity has not established yet today, we still can make
some attempts to investigate the nature of dark energy according
to some principles of quantum gravity. The holographic dark energy
model and the agegraphic dark energy  model are just such
examples, which are originated from some considerations of the
features of the quantum theory of gravity. That is to say, the
holographic and agegraphic dark energy models possess some
significant features of quantum gravity. The former, that arose a
lot of enthusiasm recently
\cite{Coh,Li,Huang,Hsu,HDE,Setare,Seta1}, is motivated from the
holographic hypothesis \cite{Suss1} and has been tested and
constrained by various astronomical observations \cite{Xin}.
However there are some difficulties in holographic dark energy
model. Choosing the event horizon of the universe as the length
scale, the holographic dark energy gives the observation value of
dark energy in the universe and can drive the universe to an
accelerated expansion phase. But an obvious drawback concerning
causality appears in this proposal. Event horizon is a global
concept of spacetime; existence of event horizon of the universe
depends on future evolution of the universe; and event horizon
exists only for universe with forever accelerated expansion. In
addition, more recently, it has been argued that this proposal
might be in contradiction to the age of some old high redshift
objects, unless a lower Hubble parameter is considered
\cite{Wei0}. The later (agegraphic dark energy) is based on the
uncertainty relation of quantum mechanics together with the
gravitational effect in general relativity. The agegraphic dark
energy model assumes that the observed dark energy comes from the
spacetime and matter field fluctuations in the universe
\cite{Cai1,Wei2,Wei1}. Since in agegraphic dark energy model the
age of the universe is chosen as the length measure, instead of
the horizon distance, the causality problem in the holographic
dark energy is avoided. The agegraphic models of dark energy  have
been examined and constrained by various astronomical observations
\cite{age,shey1,setare}.

Among the various candidates to explain the accelerated expansion,
the Chaplygin gas dark energy model has emerged as a possible
unification of dark matter and dark energy, since its cosmological
evolution is similar to an initial dust like matter and a
cosmological constant for late times. Inspired by the fact that
the Chaplygin gas possesses a negative pressure, the authors of
\cite {Gorini} have undertaken the simple task of studying a FRW
cosmology of a universe filled with this type of fluid. The
equation of state of the Chaplygin gas dark energy obeys
\cite{Kam}
\begin{eqnarray}\label{Chap}
 p_D=\frac{-A}{\rho_D},
 \end{eqnarray}
where $\rho_D>0$ and $p_D$ are, respectively, the energy density
and pressure of Chaplygin gas dark energy, and $A$ is a positive
constant. This equation of state has raised a certain interest
\cite{Baz} because of its many interesting and, in some sense,
intriguingly unique features. Some possible motivations for this
model from the field theory points of view are investigated in
\cite{Bil}. The Chaplygin gas emerges as an effective fluid
associated with D-branes \cite{Bor} and can also be obtained from
the Born-Infeld action \cite{Ben}. The connection between the
holographic models of dark energy and the Chaplygin gas energy
density has been established in \cite{seta1,seta2}.

Our aim in this Letter is to establish a correspondence between
the agegraphic dark energy scenarios and the Chaplygin gas model.
We suggest the agegraphic description of the Chaplygin gas dark
energy in FRW universe and reconstruct the potential and the
dynamics of the scalar field which describe the Chaplygin
cosmology. In the next section  we study the original agegraphic
Chaplygin gas model. In section \ref{NEW}, we establish the
correspondence between the new model of agegraphic dark energy and
the Chaplygin gas dark energy. In section \ref{Gen}, we extend our
study to the interacting agegraphic generalized Chaplygin gas dark
energy model. The last section is devoted to conclusions.

\section{THE ORIGINAL ADE as a Chaplygin gas \label{ORI}}
We assume the agegraphic Chaplygin gas dark energy is accommodated
in the Friedmann-Robertson-Walker (FRW) universe which is
described by the line element
\begin{eqnarray}
 ds^2=dt^2-a^2(t)\left(\frac{dr^2}{1-kr^2}+r^2d\Omega^2\right),\label{metric}
 \end{eqnarray}
where $a(t)$ is the scale factor, and $k$ is the curvature
parameter with $k = -1, 0, 1$ corresponding to open, flat, and
closed universes, respectively. A closed universe with a small
positive curvature ($\Omega_k\simeq0.01$) is compatible with
observations \cite{spe}. The corresponding Friedmann equation
takes the form
\begin{eqnarray}\label{Fried}
H^2+\frac{k}{a^2}=\frac{1}{3m_p^2} \left( \rho_m+\rho_D \right).
\end{eqnarray}
We introduce, as usual, the fractional energy densities such as
\begin{eqnarray}\label{Omega}
\Omega_m=\frac{\rho_m}{3m_p^2H^2}, \hspace{0.5cm}
\Omega_D=\frac{\rho_D}{3m_p^2H^2},\hspace{0.5cm}
\Omega_k=\frac{k}{H^2 a^2},
\end{eqnarray}
thus, the Friedmann equation can be written
\begin{eqnarray}\label{Fried2}
\Omega_m+\Omega_D=1+\Omega_k.
\end{eqnarray}
Inserting the equation of state (\ref{Chap}) into the relativistic
energy conservation equation, leads to a density evolving as
\begin{eqnarray}\label{rhochap}
\rho_D=\sqrt{A+\frac{B}{a^6}}.
\end{eqnarray}
where $B$ is an integration constant. We adopt the viewpoint that
the scalar field models of dark energy are effective theories of
an underlying theory of dark energy. The energy density and
pressure of the scalar field can be written as
\begin{eqnarray}\label{rhophi}
\rho_\phi=\frac{1}{2}\dot{\phi}^2+V(\phi)=\sqrt{A+\frac{B}{a^6}},\\
p_\phi=\frac{1}{2}\dot{\phi}^2-V(\phi)=\frac{-A}{\sqrt{A+\frac{B}{a^6}}},
\label{pphi}
\end{eqnarray}
Then, we can easily obtain the scalar potential and the kinetic
energy term as
\begin{eqnarray}\label{vphi}
&&V(\phi)=\frac{2Aa^6+B}{2a^6\sqrt{A+\frac{B}{a^6}}},\\
&&\dot{\phi}^2=\frac{B}{a^6\sqrt{A+\frac{B}{a^6}}}.
\label{ddotphi}
\end{eqnarray}
Now we are focussing on the reconstruction of the original
agegraphic Chaplygin gas model of dark energy. Let us first review
the origin of the agegraphic dark energy. Following the line of
quantum fluctuations of spacetime, Karolyhazy et al. \cite{Kar1}
argued that the distance $t$ in Minkowski spacetime cannot be
known to a better accuracy than $\delta{t}=\beta
t_{p}^{2/3}t^{1/3}$ where $\beta$ is a dimensionless constant of
order unity. Based on Karolyhazy relation, Maziashvili discussed
that the energy density of metric fluctuations of the Minkowski
spacetime is given by \cite{Maz}
\begin{equation}\label{rho0}
\rho_{D} \sim \frac{1}{t_{p}^2 t^2} \sim \frac{m^2_p}{t^2},
\end{equation}
where $t_{p}$ is the reduced Planck time. We use the units $c
=\hbar=k_b = 1$ throughout this Letter. Therefore one has $l_p =
t_p = 1/m_p$ with $l_p$ and $m_p$ are the reduced Planck length
and mass, respectively. The original agegraphic dark energy
density has the form (\ref{rho0}) where $t$ is chosen to be the
age of the universe
\begin{equation}
T=\int_0^a{\frac{da}{Ha}},
\end{equation}
Thus, the energy density of the original agegraphic dark energy is
given by \cite{Cai1}
\begin{equation}\label{rho1}
\rho_{D}= \frac{3n^2 m_{p}^2}{T^2},
\end{equation}
where the numerical factor $3n^2$ is introduced to parameterize
some uncertainties, such as the species of quantum fields in the
universe, the effect of curved space-time (since the energy
density is derived for Minkowski space-time), and so on. The dark
energy density (\ref{rho1}) has the same form as the holographic
dark energy, but  the length measure is chosen to be the age of
the universe instead of the horizon radius of the universe. Thus
the causality problem in the holographic dark energy is avoided.
Using Eqs. (\ref{Omega}) and (\ref{rho1}), we have
\begin{eqnarray}\label{Omegaq}
\Omega_D=\frac{n^2}{H^2T^2}.
\end{eqnarray}
We assume the agegraphic dark energy and dark matter evolve
according to their conservation laws
\begin{eqnarray}
&&\dot{\rho}_D+3H\rho_D(1+w_D)=0,\label{consq}\\
&&\dot{\rho}_m+3H\rho_m=0, \label{consm}
\end{eqnarray}
where $w_D$ is the equation of state parameter of agegraphic dark
energy. Taking the  derivative with respect to the cosmic time of
Eq. (\ref{rho1})  and using Eq. (\ref{Omegaq}) we get
\begin{eqnarray}\label{rhodot}
\dot{\rho}_D=-2H\frac{\sqrt{\Omega_D}}{n}\rho_D.
\end{eqnarray}
Inserting this relation into Eq. (\ref{consq}), we obtain the
equation of state parameter of the original agegraphic dark energy
\begin{eqnarray}\label{wq}
w_D=-1+\frac{2}{3n}\sqrt{\Omega_D}.
\end{eqnarray}
Differentiating Eq. (\ref{Omegaq}) and using relation
${\dot{\Omega}_D}={\Omega'_D}H$, we reach
\begin{eqnarray}\label{Omegaq2}
{\Omega'_D}=\Omega_D\left(-2\frac{\dot{H}}{H^2}-\frac{2}{n
}\sqrt{\Omega_D}\right),
\end{eqnarray}
where the dot is the derivative with respect to the cosmic time
and the prime denotes the derivative with respect to $x=\ln{a}$.
Taking the derivative of  both side of the Friedman equation
(\ref{Fried}) with respect to the cosmic time, and using Eqs.
(\ref{Fried2}), (\ref{rho1}), (\ref{Omegaq}) and (\ref{consm}), it
is easy to find
\begin{eqnarray}\label{Hdot}
\frac{\dot{H}}{H^2}=-\frac{3}{2}(1-\Omega_D)-\frac{\Omega^{3/2}_D}{n}-\frac{\Omega_k}{2}.
\end{eqnarray}
Substituting this relation into Eq. (\ref{Omegaq2}), we obtain the
equation of motion of agegraphic dark energy
\begin{eqnarray}\label{Omegaq3}
{\Omega'_D}&=&\Omega_D\left[(1-\Omega_D)\left(3-\frac{2}{n}\sqrt{\Omega_D}\right)
+\Omega_k\right].
\end{eqnarray}
Next, we establish the connection between the original agegraphic
dark energy and Chaplygin gas energy density. Combining Eqs.
(\ref{rhochap}) and (\ref{rho1}), we obtain
\begin{eqnarray}\label{BA1}
B=a^6\left(9n^4m^4_pT^{-4}-A\right).
\end{eqnarray}
Using Eqs. (\ref{Chap}), (\ref{rhochap}) and (\ref{wq}) one can
write
\begin{eqnarray}\label{BA2}
w_D=\frac{-A}{\rho_D^2}=\frac{-A}{A+\frac{B}{a^6}}=-1+\frac{2}{3n}\sqrt{\Omega_D}.
\end{eqnarray}
Substituting $B$ in the above equation, we obtain following
relation for $A$:
\begin{eqnarray}\label{A}
A=9n^4m^4_pT^{-4}\left(1-\frac{2}{3n}\sqrt{\Omega_D}\right).
\end{eqnarray}
Therefore the constant $B$ is given by
\begin{eqnarray}\label{B}
B=6a^6n^4m^4_pT^{-4}\frac{\sqrt{\Omega_D}}{n}.
\end{eqnarray}
Finally, we rewrite the scalar potential and kinetic energy term
as
\begin{eqnarray}\label{vphi2}
V(\phi)&=&n^2m^2_pT^{-2} \left(3-\frac{\sqrt{\Omega_D}}{n}\right) =m^2_pH^2 \Omega_D\left(3-\frac{\sqrt{\Omega_D}}{n}\right),\\
\dot{\phi}&=&n m_p  T^{-1} \sqrt{\frac{2}{n}\Omega^{1/2}_D}=m_pH
\sqrt{\frac{2}{n}{\Omega^{3/2}_D}}.\label{dotphi2}
\end{eqnarray}
Using relation $\dot{\phi}=H{\phi'}$, we get
\begin{eqnarray}\label{primephi}
{\phi'}&=&m_p \sqrt{\frac{2}{n}{\Omega^{3/2}_D}}.
\end{eqnarray}
Consequently, we can easily obtain the evolution behavior of the
scalar field
\begin{eqnarray}\label{phi}
\phi(a)-\phi(a_0)=\int_{a_0}^{a}{\frac {m_p}{a}
\sqrt{\frac{2}{n}{\Omega^{3/2}_D}}da},
\end{eqnarray}
where $a_0$  is the  present value of the scale factor, and
$\Omega_D$ can be obtained through Eq. (\ref{Omegaq3}).
\section{THE NEW ADE as a Chaplygin gas \label{NEW}}
Soon after the original agegraphic dark energy model was
introduced by Cai \cite{Cai1}, a new model of agegraphic dark
energy  was proposed in \cite{Wei2}, while the time scale is
chosen to be the conformal time $\eta$ instead of the age of the
universe. This new agegraphic dark energy  contains some new
features different from the original agegraphic dark energy and
overcome some unsatisfactory points. For instance, the original
agegraphic dark energy suffers from the difficulty to describe the
matter-dominated epoch while the new agegraphic dark energy
resolved this issue \cite{Wei2}. The energy density of the new
agegraphic dark energy can be written
\begin{equation}\label{rho1new}
\rho_{D}= \frac{3n^2 m_{p}^2}{\eta^2},
\end{equation}
where the conformal time is given by
\begin{equation}
\eta=\int{\frac{dt}{a}}=\int_0^a{\frac{da}{Ha^2}}.
\end{equation}
The fractional energy density of the new agegraphic dark energy is
now given by
\begin{eqnarray}\label{Omegaqnew}
\Omega_D=\frac{n^2}{H^2\eta^2}.
\end{eqnarray}
Taking the  derivative with respect to the cosmic time of Eq.
(\ref{rho1new}) and using Eq. (\ref{Omegaqnew}) we get
\begin{eqnarray}\label{rhodotnew}
\dot{\rho}_D=-2H\frac{\sqrt{\Omega_D}}{na}\rho_D.
\end{eqnarray}
Inserting this relation into Eq. (\ref{consq}) we obtain the
equation of state parameter of the new agegraphic dark energy
\begin{eqnarray}\label{wqnew}
w_D=-1+\frac{2}{3na}\sqrt{\Omega_D}.
\end{eqnarray}
Then we obtain, following the approach of the previous section,
the evolution behavior of the new agegraphic dark energy,
\begin{eqnarray}\label{Omegaq3new}
{\Omega'_D}&=&\Omega_D\left[(1-\Omega_D)\left(3-\frac{2}{na}\sqrt{\Omega_D}\right)
+\Omega_k\right].
\end{eqnarray}
Next, we construct the new agegraphic Chaplygin gas model,
connecting the Chaplygin gas model with new agegraphic dark
energy. Identifying Eq. (\ref{rhochap}) with (\ref{rho1new}) we
have
\begin{eqnarray}\label{BA}
B=a^6\left(9n^4m^4_p\eta^{-4}-A\right).
\end{eqnarray}
Using Eqs. (\ref{Chap}), (\ref{rhochap}) and (\ref{wqnew}), we
reach
\begin{eqnarray}\label{ABn}
w_D=\frac{-A}{A+\frac{B}{a^6}}=-1+\frac{2}{3na}\sqrt{\Omega_D}.
\end{eqnarray}
Combining this equation with Eq. (\ref{BA}), we obtain
\begin{eqnarray}\label{An}
A&=&9n^4m^4_p\eta^{-4}\left(1-\frac{2}{3na}\sqrt{\Omega_D}\right),\\
B&=&6a^6n^4m^4_p\eta^{-4}\frac{\sqrt{\Omega_D}}{na}\label{Bn}.
\end{eqnarray}
Finally, we reconstruct the scalar potential and kinetic energy
term as
\begin{eqnarray}\label{vphi2new}
V(\phi)&=&n^2m^2_p\eta^{-2}
\left(3-\frac{\sqrt{\Omega_D}}{na}\right)=m^2_pH^2
\Omega_D\left(3-\frac{\sqrt{\Omega_D}}{na}\right),
\\ \dot{\phi}&=&n m_p  \eta^{-1}
\sqrt{\frac{2}{na}{\Omega^{1/2}_D}}=m_pH
\sqrt{\frac{2}{na}{\Omega^{3/2}_D}}.\label{dotphi2new}
\end{eqnarray}
Eq. (\ref{dotphi2new}) can also be reexpressed as
\begin{eqnarray}\label{primephinew}
{\phi'}&=&m_p \sqrt{\frac{2}{na}{\Omega^{3/2}_D}}.
\end{eqnarray}
Therefore, we can obtain the evolutionary form of the scalar field
\begin{eqnarray}\label{phinew}
\phi(a)-\phi(a_0)=\int_{a_0}^{a}{\frac{m_p}{a}
\sqrt{\frac{2}{na}{\Omega^{3/2}_D}}da},
\end{eqnarray}
where $\Omega_D$ can be derived from Eq. (\ref{Omegaq3new}).
\section{Interacting new agegraphic generalized Chaplygin gas\label{Gen}}
In this section we extend our study to the generalized Chaplygin
gas when there is an interaction between generalized Chaplygin gas
energy density and dark matter. The total energy density is
$\rho=\rho_{m}+\rho_{D}$, where $\rho_{m}$ and $\rho_{D}$ are the
energy density of dark matter and dark energy, respectively. The
total energy density satisfies a conservation law
\begin{equation}\label{cons}
\dot{\rho}+3H(\rho+p)=0.
\end{equation}
However, since we consider the interaction between dark matter and
dark energy, $\rho_{m}$ and $\rho_{D}$ do not conserve separately;
they must rather enter the energy balances
\begin{eqnarray}
&&\dot{\rho}_m+3H\rho_m=Q, \label{consm2}
\\&& \dot{\rho}_D+3H\rho_D(1+w_D)=-Q,\label{consq2}
\end{eqnarray}
where $Q$ denotes the interaction term and can be taken as $Q
=3b^2 H\rho$  with $b^2$  being a coupling constant. In the
generalized Chaplygin gas approach \cite{Ben}, the equation of
state (\ref{Chap}) is generalized to
\begin{eqnarray}\label{Chap2}
p_D=\frac{-A}{{\rho_D^{\alpha}}}.
 \end{eqnarray}
 The above equation of state leads to a density evolution as
\begin{eqnarray}\label{rhochap2}
\rho_D=\left({A+{B}a^{-3\beta}}\right)^{1/\beta},
\end{eqnarray}
where $\beta=\alpha+1$. Thus we have
\begin{eqnarray}\label{wchap2}
w_D=\frac{p_D}{\rho_D}=\frac{-A}{{A+{B}a^{-3\beta}}}.
\end{eqnarray}
Inserting  relation (\ref{rhodotnew}) into Eq. (\ref{consq2}) and
using Eqs. (\ref{Omega}) and  (\ref{Fried2}) , we obtain the
equation of state parameter of the interacting new agegraphic dark
energy
\begin{eqnarray}\label{wqInt}
w_D=-1+\frac{2}{3na}\sqrt{\Omega_D}-\frac{b^2}{\Omega_D}
(1+\Omega_k).
\end{eqnarray}
The evolution behavior of the new agegraphic dark energy is given
by
\begin{eqnarray}\label{Omegaq3newInt}
{\Omega'_D}&=&\Omega_D\left[(1-\Omega_D)\left(3-\frac{2}{na}\sqrt{\Omega_D}\right)
-3b^2(1+\Omega_k)+\Omega_k\right].
\end{eqnarray}
We now establish the correspondence between the new agegraphic
dark energy and generalized Chaplygin gas energy density.
Identifying Eq. (\ref{rho1new}) with  Eq. (\ref{rhochap2}) and
using Eq. (\ref{Omegaqnew}) we get
\begin{eqnarray}\label{rhochap3}
\left(3m^2_pH^2\Omega_D\right)^{\beta}={A+{B}a^{-3\beta}}.
\end{eqnarray}
Combining Eqs. (\ref{wchap2}) and (\ref{wqInt}) we find
\begin{eqnarray}\label{rhochap4}
A=-\left({A+{B}a^{-3\beta}}\right)\left[-1+\frac{2}{3na}\sqrt{\Omega_D}-\frac{b^2}{\Omega_D}
(1+\Omega_k)\right].
\end{eqnarray}
Solving Eqs. (\ref{rhochap3}) and (\ref{rhochap4}) we obtain
\begin{eqnarray}\label{Aint}
A&=&\left(3m^2_pH^2\Omega_D\right)^{\beta}\left[1-\frac{2}{3na}\sqrt{\Omega_D}+\frac{b^2}{\Omega_D}
(1+\Omega_k)\right],\\
B&=&\left(3m^2_pH^2\Omega_Da^3\right)^{\beta}\left[\frac{2}{3na}\sqrt{\Omega_D}-\frac{b^2}{\Omega_D}
(1+\Omega_k)\right].\label{Bint}
\end{eqnarray}
Next we regard the scalar field model as an effective description
of an underlying theory of dark energy with energy density and
pressure
\begin{eqnarray}\label{rhophiI}
\rho_\phi&=&\frac{1}{2}\dot{\phi}^2+V(\phi)=\left({A+{B}a^{-3\beta}}\right)^{1/\beta},\\
p_\phi&=&\frac{1}{2}\dot{\phi}^2-V(\phi)=-A\left({A+{B}a^{-3\beta}}\right)^{-\alpha/\beta},
\label{pphiI}
\end{eqnarray}
where we have identified  $\rho_\phi$ with $\rho_D$. Substituting
$A$ and $B$ into Eqs. (\ref{rhophiI})  and (\ref{pphiI}) one can
easily find the scalar potential and the kinetic energy term as
\begin{eqnarray}\label{vphi2I}
V(\phi)&=&m^2_pH^2 \Omega_D\left(3-\frac{\sqrt{\Omega_D}}{na}+\frac{3b^2}{2}\frac{(1+\Omega_k)}{\Omega_D}\right),\\
\dot{\phi}&=&m_pH\left(
\frac{2}{na}{\Omega^{3/2}_D}-3b^2(1+\Omega_k)\right)^{1/2}.\label{dotphi2I}
\end{eqnarray}
Using Eq. (\ref{dotphi2I}), Eq. (\ref{vphi2I}) can be reexpressed
as
\begin{eqnarray}\label{vphiphi}
V(\phi)&=&3m^2_pH^2 \Omega_D\left(1-\frac{\dot{\phi}^2}{6m^2_pH^2
\Omega_D}\right).
\end{eqnarray}
We can also rewrite Eq. (\ref{dotphi2I}) as
\begin{eqnarray}\label{primephiI}
{\phi'}&=&m_p\left(
\frac{2}{na}{\Omega^{3/2}_D}-3b^2(1+\Omega_k)\right)^{1/2}.
\end{eqnarray}
Consequently, we can easily obtain the evolutionary form of the
field by integrating the above equation. The result is
\begin{eqnarray}\label{phiI}
\phi(a)-\phi(a_0)=\int_{a_0}^{a}{\frac {m_p}{a}\sqrt{
\frac{2}{na}{\Omega^{3/2}_D}-3b^2(1+\Omega_k)}da},
\end{eqnarray}
where $\Omega_D$ is given by Eq. (\ref{Omegaq3newInt}). In this
way we connect the interacting new agegraphic dark energy with a
generalized Chaplygin gas energy density and reconstruct the
potential of the agegraphic Chaplygin gas.
\section{Conclusions\label{CONC}}
In this Letter, we have established a correspondence between the
agegraphic dark energy scenarios and the Chaplygin gas model of
dark energy in non-flat FRW cosmology. A so-called agegraphic dark
energy model has been proposed recently by Cai \cite{Cai1}, based
on the uncertainty relation of quantum mechanics together with the
gravitational effect in general relativity. Since the original
agegraphic dark energy model suffers from the difficulty to
describe the matter-dominated epoch, a new model of agegraphic
dark energy was proposed in \cite{Wei2} while the time scale is
chosen to be the conformal time $\eta$ instead of the age of the
universe. We have adopted the viewpoint that the scalar field
models of dark energy are effective theories of an underlying
theory of dark energy. If we regard the scalar field model as an
effective description of such a theory, we should be capable of
using the scalar field model to mimic the evolving behavior of the
agegraphic dark energy and reconstructing this scalar field model
according to the evolutionary behavior of agegraphic dark energy.
We have reconstructed the potential of the agegraphic scalar field
as well as the dynamics of the scalar field which describe the
Chaplygin cosmology. Finally, we have extended our study to the
agegraphic generalized Chaplygin gas dark energy model when there
is an interaction between generalized Chaplygin gas energy density
and dark matter.

\acknowledgments{This work has been supported by Research
Institute for Astronomy and Astrophysics of Maragha.}

\end{document}